\documentclass[prl,reprint,showpacs,floatfix]{revtex4-1}

\usepackage{graphicx}
\usepackage{dcolumn}
\usepackage{bm}
\usepackage{color}
\usepackage{url}
\usepackage{amsmath}
\usepackage{amssymb}
\usepackage{mathrsfs}

\begin{document}

\title{KATRIN Sensitivity to Sterile Neutrino Mass\\ In The Shadow of Lightest Neutrino Mass}

\author{Arman Esmaili}
\email{aesmaili@ifi.unicamp.br}
\affiliation{Instituto de F\'isica Gleb Wataghin - UNICAMP, 13083-859, Campinas, SP, Brazil\ }
\author{Orlando L. G. Peres}
\email{orlando@ifi.unicamp.br}
\affiliation{Instituto de F\'isica Gleb Wataghin - UNICAMP, 13083-859, Campinas, SP, Brazil\ }
\affiliation{Arizona  State University, Tempe, AZ 85287-1504}
\affiliation{The Abdus Salam International Centre for Theoretical Physics\\
Strada Costiera 11, I-34014 Trieste, Italy}

\begin{abstract}
The presence of light sterile neutrinos would strongly modify the
energy spectrum of the Tritium $\beta$-electrons. We perform an
analysis of the KATRIN experiment's sensitivity by scanning almost all
the allowed region of neutrino mass-squared difference and mixing
angles of the 3+1 scenario. We consider the effect of the unknown
absolute mass scale of active neutrinos on the sensitivity of KATRIN
to the sterile neutrino mass.  We show that after 3 years of
data-taking, the KATRIN experiment can be sensitive to mixing angles as
small as $\sin^2 2\theta_s\sim 10^{-2}$. Particularly we show that for
small mixing angles, $\sin^2 2\theta_s \lesssim0.1$, the KATRIN
experiment can gives the strongest limit on active-sterile mass-squared
difference.
\end{abstract}

\maketitle

\section{\label{sec:intro}Introduction}

There is a consensus that nonzero masses of neutrinos and the
nontrivial mixing between them is the plausible framework to explain
the outstanding results of plenty of neutrino oscillation
experiments. The standard approach is to have a three active neutrino
scenario, with at least two nonzero masses and two reasonably large
mixing angles~\cite{GonzalezGarcia:2007ib}.

An interesting extension to this standard scenario is the existence of
extra light sterile neutrino states which arose for the first time in
the light of LSND experiment~\cite{Aguilar:2001ty} showing evidence
for $\bar{\nu}_{\mu} \rightarrow \bar{\nu}_{e}$ oscillation (see also
the recent MiniBooNE data~\cite{AguilarArevalo:2010wv} that seems to
corroborate). Another evidence of the presence of light sterile
neutrinos is the so-called reactor neutrino
anomaly~\cite{Mention:2011rk}. This anomaly is the departure from
unity of the ratio of observed rate of events to the predicted rate in very short-baseline
reactor neutrino experiments. This departure prompted by the re-evaluation of $\bar{\nu}_{e}$ reactor flux
that revealed 2.5 \% increase in the flux~\cite{Mention:2011rk}. Also
the Gallium anomaly~\cite{Giunti:2010zu} show a deficit of $\nu_e$
produced by intense radioactive sources, such that the ratio of observed rate
to the predicted rate is
$0.86\pm0.05$~\cite{Giunti:2010zu}. All these anomalies can be understood by adding one (or more) light
sterile neutrino states to the scenario with three active neutrinos, with active-sterile mass-squared difference $\Delta{m}^2_{\text{SBL}}
\gtrsim 0.1 \, \text{eV}^2$.
The 3+1 scenario~\cite{Barger:1998bn,allowed}
is the simplest scenario to accommodate the presence of light sterile
neutrinos. The model is composed of 4 flavor states $\nu_{\alpha}$,
$\alpha=e,\mu,\tau$ and $s$, that are the mixture of mass eigenstates
$\nu_i$ with masses $m_i$, $i=1,2,3,4$. We associate the mass scale
that induce the very short-baseline oscillations,
$\Delta{m}^2_{\text{SBL}}$, with the mass difference $\Delta{m}^2_{41}\equiv m_4^2-m_1^2$.
It should be noticed that in order to explain the reactor anomaly, a nonzero component of $\nu_4$ in
the state $\nu_e$ is necessary.

A new generation of tritium beta decay
experiments, like the KArlsruhe TRItium
Neutrino (KATRIN) experiment~\cite{katrin}, was proposed to search
kinematically for the neutrino mass by measuring the energy spectrum
of the electrons from the beta decay of tritium
${}^3$H$\to{}^3$He${}^+ + e^- + \bar{\nu}_e$ (see also other
proposals~\cite{other1}).
A nonzero neutrino mass results in displacement of the endpoint energy
in electron spectrum which is the focus region to probe in KATRIN.
In this Letter we analyze the capability of KATRIN experiment in the
search for endpoint displacement in the energy spectrum
of $\beta$-electrons due to the lightest neutrino mass $m_1$ and the
irregularities in the shape of energy spectrum due to the heavier (mostly sterile)
neutrino mass $m_4$ and its nonzero mixing with electron neutrino. We
discuss for the first time the role of the lightest neutrino mass
$m_1$ in determination of the sensitivity of KATRIN experiment to the
oscillation parameter $\Delta m_{41}^2$. We show
that with the present KATRIN design, this experiment is the only one
sensitive to the part of parameter space corresponding to very small
active-sterile mixing angle and large active-sterile mass-squared difference.

\section{\label{sec:beta}Tritium Beta Decay at KATRIN}
The KATRIN experiment~\cite{katrin}
have the following setup.
Injected molecular tritium gas at Tritium
Laboratory Karlsruhe provides high luminosity $\beta$-electrons
emitting isotropically. The electrons will be guided by the gradient
of a magnetic field to the so-called MAC-E-Filter (Magnetic Adiabatic
Collimation combined with an Electrostatic Filter) spectrometer. The
ratio of the minimum magnetic field at the central plane of the spectrometer ($B_A=3\times
10^{-4}$~T) to the maximum magnetic field near the
tritium source ($B_{{\rm max}}=6$~T) determines the relative sharpness
of the energy filtering of MAC-E-Filter.
Also applying a magnetic field $B_S=3.6$~T at the tritium source suppresses the entrance
of electrons with large initial emission angle.
In the spectrometer, with the help of an electric field parallel to the electron's propagation path, it
is possible to make an electrostatic barrier $qU$ which can be passed
just by electrons with energy higher than the height of the
barrier. Taking all together, the transmission function of the KATRIN
spectrometer as a function of electron kinetic energy $K_e$ and
retarding potential $qU$ is
\begin{displaymath}
T(K_e,qU)=\left\{
     \begin{array}{lcc}
      0 & {\rm if} &
      K_e-qU<0\\ \frac{1-\sqrt{1-\frac{K_e-qU}{K_e}\frac{B_S}{B_A}}}{1-\sqrt{1-\frac{\Delta
            K_e}{K_e}\frac{B_S}{B_A}}} & {\rm if} & 0\le K_e-qU\le
      \Delta K_e\\ 1 & {\rm if} & K_e-qU > \Delta K_e
     \end{array}
   \right.
\end{displaymath} 
where $\Delta K_e/K_e=B_A/B_{{\rm max}}$ is the relative sharpness of
the filter. However, electrons can undergo inelastic scattering with
tritium molecules in the source which can change the spectrum. Taking
into account the probability of multiple inelastic scattering, the
transmission function modifies to the following convoluted form
\begin{eqnarray}
T^\prime (K_e,qU) & &\;\,= \int_0^{K_e-qU} T(K_e-\epsilon,qU)
\\ &&\!\!\!\!\!\times \left[P_0 \delta(\epsilon)+P_1 f(\epsilon) +P_2
  (f\otimes f)(\epsilon) + \ldots\right] {\rm d}\epsilon ,\nonumber
\end{eqnarray}
where $P_n$ is the probability that the electron scatters $n$ times
off the tritium molecules before leaving the source and $f(\epsilon)$ is
the energy loss function at each scattering~\cite{ffunc}. The symbol $\otimes$ defines the
following convolution:
\begin{equation*}
(f\otimes f) (\epsilon) = \int_{0}^{K_e-qU} f(\epsilon^\prime)
  f(\epsilon -\epsilon^\prime) {\rm d}\epsilon^\prime .
\end{equation*}

The rate of the electrons passing the potential barrier $qU$ and arriving
at the detector is
\begin{eqnarray}
&\!\!\!\!\!\!\!\!\!\!\!\!\!\!\!\!\!\!\!\!\!\!\!\!\!\!\!\!\!\!\!\!\!\!\!\!\!\!\!\!\!\!\!\!\!\!\!\!\!\!\!\!\!\!\!\!\!\!\!\!\!\!\!\!\!\!\!\!\!\!\!\!\!\!\!\!\!\!\!\!\!\!\!\!\!\!\!\!\!S(Q,qU,[U_{ei}],[m_\nu])=\nonumber\\ 
&\qquad\qquad\!\!\!\!\int_0^\infty
  \beta (K_e,Q,[U_{ei}],[m_\nu]) T^\prime (K_e,qU)\, {\rm d}K_e \, ,
\label{spectrum}
\end{eqnarray}
where the symbols $[m_\nu]=\{m_1,\ldots,m_n\}$ and
$[U_{ei}]=\{U_{e1},\ldots,U_{en} \}$ denotes respectively the set of
the masses of neutrino mass eigenstates $\nu_i$ and the elements of
the first row of Pontecorvo–-Maki–-Nakagawa-–Sakata (PMNS) $U_{{\rm
    PMNS}}$~\cite{Pontecorvo:1957cp} mixing matrix. The function
$\beta$ gives the spectrum of electrons in beta decay
\begin{eqnarray} \label{beta} 
\beta
(K_e,Q,[U_{ei}],[m_\nu])= \\
&\!\!\!\!\!\!\!\!\!\!\!\!\!\!\!\!\!\!\!\!\!\!\!\!\!\!\!\!\!\!\!\!\!\!\!\!\!\!\!\!\!\!\!\!\!\!\!\!\!\!\!\!\!\!\!\!\!\!\!\!\!\!\!\!\!
N_s F(Z,K_e) E_e p_e\displaystyle\sum_{i,j}\left[ p_i {\cal E}_i
  |U_{ej}|^2\sqrt{{\cal E}_i^2-m_j^2}\Theta({\cal
    E}_i-m_j)\right],\nonumber
\end{eqnarray}
where ${\cal E}_i=Q-W_i-K_e$. In the above equation $E_e$ and $p_e$
are respectively the electron's energy and momentum; $F(Z,K_e)$ is the
Fermi function which takes into account the electrostatic interaction
of the emitted electron with the daughter nucleus with
$Z=2$~\cite{fermifunc}; $W_i$ and $p_i$ are respectively the
excitation energy and transition probability for the excited state $i$
of the daughter nucleus~\cite{excite}; and the Heaviside step function $\Theta$ guarantees the conservation of energy. The index $i$ runs over the
excited states and index $j$ runs over the neutrino mass
eigenstates. The factor $N_s$ determines the total number of emitted
electrons which for the KATRIN design parameters is $1.47\times
10^{-13}\,{\rm s}^{-1}\,{\rm eV}^{-5}$~\cite{katrin}.

\section{\label{sec:analysis}Sensitivity of KATRIN to Sterile Neutrino}

\begin{figure}[h!]
\includegraphics[scale=0.5]{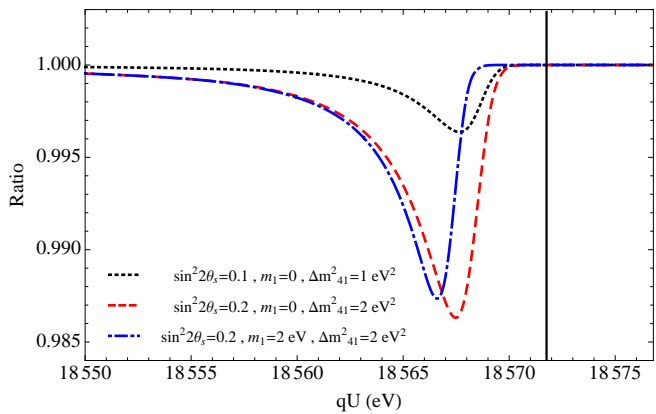}
\caption{\label{ratios} The ratios of the total rate in Eq.~(\ref{rationeq}) for various mixing and mass parameters. The vertical line shows the $Q=18571.8$~eV.}
\end{figure}

\begin{figure}[t]
\includegraphics[scale=0.47]{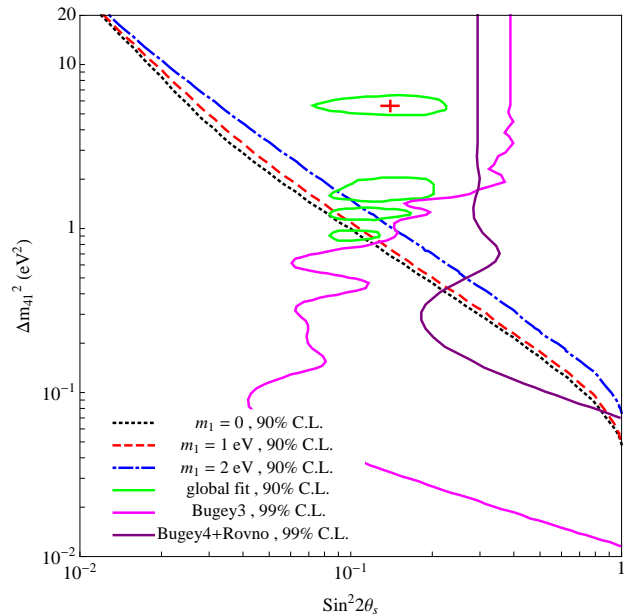}
\caption{\label{sen} 
The $90\%$~C.L. contours of the KATRIN experiment in the $(\sin^2
2\theta_s,\Delta m_{41}^2)$ plane. The black (dotted), red (dashed)
and blue (dot-dashed) curves correspond respectively to
$m_1=0,1,2$~eV. 
The green (solid) curves show the $90\%$~C.L. allowed
region and the red cross the best fit point for the global fit of the
data for the 3+1 scheme~\cite{allowed}. The magenta and purple curves
show respectively the Bugey3 and Bugey4+Rovno exclusion
curves~\cite{Giunti:2010zu}.  
}
\end{figure}

The functional form of the $\beta$-electron spectrum in
Eq.~(\ref{beta}) depends on the set of masses
$[m_\nu]=\{m_1,\ldots,m_n\}$. As discussed
in~\cite{Farzan:2001cj}
, for the case that the energy
resolution of the experiment near the endpoint and the energy
interval that is probed by the experiment is much larger than the mass
splittings, it is possible to replace the set of masses $[m_\nu]$ with
an effective mass $m_\beta \equiv \sqrt{\sum_i m_i^2 |U_{ei}|^2}$.
However, in the case of sterile neutrino
with a mass-squared difference $\sim 1\,{\rm eV}^2$, this
approximation fails and the error of using effective mass in the fit
of spectrum becomes large. Here we use the exact form of
Eq.~(\ref{beta}) with four mass parameters $\{m_1,m_2,m_3,m_4\}$ in
the case of 3+1 scheme. However, from these four masses just $m_1$ and
$m_4$ enter the analysis and the other two are fixed by $m_2=\sqrt{m_1^2+\Delta m_{21}^2}$ and $m_3=\sqrt{m_1^2+\Delta m_{31}^2}$;
where the values of $\Delta m_{21}^2$ and $\Delta m_{31}^2$ are fixed
by oscillation phenomenology~\cite{GonzalezGarcia:2007ib}.
For the elements of the PMNS mixing matrix we use such
parameterization of the $U_{4\times 4}$ that its $3\times 3$
sub-matrix for the light active masses reduces to the PDG
parameterization~\cite{pdg}. Thus, the element $U_{e4}$ just depends
on one mixing angle $\theta_s$, through $|U_{e4}|=\sin\theta_s$.
With the above mentioned considerations, the rate of the events in
Eq.~(\ref{spectrum}) is a function of parameters
$(Q,qU,U_{e4},m_1,m_4)$. The total rate is the sum
of signal rate $S$ in Eq.~(\ref{spectrum}) and the expected rate of
the background events $N_b$ which for the KATRIN is 10 mHz~\cite{katrin}. 
For the $Q$-value of the tritium beta decay we use the central 
value of a recent measurement $Q=18571.8\pm1.2 \; {\rm eV}$~\cite{qvalue}.
To illustrate the behavior of sterile admixture, we
define the following ratio 
\begin{equation}
\frac{S(\sin^2 2\theta_s,m_1,\Delta m^2_{41})+N_b}{S(\sin^2
  2\theta_s=0,m_1,\Delta m^2_{41}=0)+N_b}~,
\label{rationeq}
\end{equation}
where $\sin^2 2\theta_s=4|U_{e4}|^2(1-|U_{e4}|^2)$.
We plotted this ratio in Fig.~\ref{ratios} for 
different set of the parameters $\sin^2 2\theta_s$, $m_1$ and $\Delta m^2_{41}$.
Comparing the black (dotted) curve with the red (dashed) curve in Fig.~\ref{ratios}, it is easy to see the change in $\beta$-spectrum for different values of the  
mass $m_4$ for a vanishing light mass $m_1=0$. The height of minimum depends on the values of $\sin^2 2\theta_s$ and
$\Delta m^2_{41}$ and the position of the minimum depends only on $m_4$. 
However, in the comparison between the red (dashed) and blue (dot-dashed) curves in
Fig.~\ref{ratios}, we see that by the inclusion of a nonzero value for 
$m_1$ the two curves with the same $\Delta m^2_{41}$ cross each other, implying that the lack of knowledge about the value of
$m_1$ can lay a shadow on the determination of $\Delta m^2_{41}$.
To quantify the sensitivity of the KATRIN experiment to the sterile
neutrino mass we define the following $\chi^2$ function
\begin{eqnarray}\label{chi2}
\chi^2 (Q,U_{e4},m_1,m_4,R_s,R_b)= & \\
&\!\!\!\!\!\!\!\!\!\!\!\!\!\!\!\!\!\!\!\!\!\!\!\!\!\!\!\!\!\!\!\!\!\!\!\!\!\!\!\!\!\!\!\!\!\!\!\!\!\!\!\!\!\!\!\!\!\!\!\!\!\!\!\!\!\!\!\!\!\!\!\!\!\!\!\!\!\displaystyle\sum_i \frac{\left(N_{{\rm
      exp}}([qU]_i)-N_{{\rm
      th}}(Q,[qU]_i,U_{e4},m_1,m_4,R_s,R_b)\right)^2}{\sigma^2}~,  \nonumber
\end{eqnarray}
where $\sigma=\sqrt{N_{{\rm exp}}}$ is the statistical standard
deviation. In Eq.~(\ref{chi2}) the $N_{{\rm th}}$ corresponds to the
number of detected $\beta$-electrons when the retarding potential has
the value $[qU]_i$ which is calculated by multiplying the rate in
Eq.~(\ref{spectrum}) by the time spent for the measurement:
\begin{equation}
N_{{\rm th}}(\ldots,[qU]_i,\ldots)= t[i] \cdot \left(R_s
S(\ldots,[qU]_i,\ldots)+R_b N_b\right)~,
\nonumber
\end{equation}
where $R_s$ and $R_b$ are respectively the normalization factors of
signal and background events; and $N_b=10$~mHz. The $N_{{\rm exp}}$
denotes the experimental number of events assuming $m_4$ and $U_{e4}$
equal to zero. The index $i$ in Eq.~(\ref{chi2}) runs in 31 steps such
that the retarding potential covers the range 
$qU \in [Q-20~{\rm eV},Q+5~{\rm eV}]$. For the time 
$t[i]$ spent at each step of the
retarding potential we use the optimized measurement time proposed by
the KATRIN collaboration (see Figure 131 in Ref.~\cite{katrin}). We
minimize the $\chi^2$ function with respect to the normalization
factors $R_s$ and $R_b$ analytically and with respect to $Q$
numerically in its uncertainty range. Fig.~\ref{sen} shows the
$90\%$~C.L. sensitivity contours of the KATRIN experiment in the
$(\sin^2 2\theta_s,\Delta m_{41}^2)$ plane for the total measurement
time $\sum_i t[i]=3$~years. The black (dotted), red (dashed) and blue
(dot-dashed) curves correspond respectively to $m_1=0,1,2$~eV. The
green (solid) curves show the $90\%$~C.L. allowed region from the
global fit of the short-baseline oscillation data~\cite{allowed} and
the red cross shows the best-fit value.

We can conclude from Fig.~\ref{sen} that after three years of
data-taking, the KATRIN experiment can exclude the main part of the
current allowed region of 3+1 scenario. Beside that, the KATRIN is the
most sensitive experiment in the upper left part of $(\sin^2
2\theta_s,\Delta m_{41}^2)$ plane. As can be seen, the experiment is
sensitive to mixing angles as small as $\sin^2
2\theta_s\sim10^{-2}$. For comparison we have shown in Fig.~\ref{sen}
the present exclusion curves of Bugey3 (magenta solid curve) and
Bugey4+Rovno (purple solid curve)~\cite{Giunti:2010zu}.

We have also found that, for the first time, varying the value of
lightest neutrino mass $m_1$ can affect the sensitivity to the large mass
$m_4$.  For example, for a fixed value of mixing angle $\sin^2
2\theta_s=0.1$, the sensitivity of the experiment is $\Delta
m_{41}^2=0.98, 1.1~{\rm and}~ 1.5 ~{\rm eV}^2$ for respectively
$m_1=0,1~{\rm and}~2$~eV as can be seen from Fig.~\ref{sen}. This
implies a correlation between the discovery potential of $\Delta m_{41}^2$
and the value of $m_1$. For smaller (larger) values of $m_1$ the correlation is weaker
(stronger), and also the correlation depends on the mixing angle. For
very small mixing angles the correlation disappear
because of the weak $m_4$ contribution, and for $\sin^2 2\theta_s=1$
case, which corresponds to equal admixture of $\nu_1$ and $\nu_4$ in
$\nu_e$, the correlation still exists. Also, it should be noticed that the large values of $m_1$ are in potential conflict with the standard $\Lambda$CDM model of cosmology, although extensions to non-minimal cosmological models can reduce the conflict~\cite{Hamann:2011ge}.

\section{\label{sec:comparation}Comparison with earlier works}

The effect of light sterile neutrinos in beta decay experiments was formerly
discussed in Refs.~\cite{Farzan:2001cj,PhysRevD75013003,Riis:2010zm,Formaggio:2011jg}. Our
results are in agreement with the estimations of~\cite{Farzan:2001cj,PhysRevD75013003}. 
In Ref.~\cite{Riis:2010zm} a general analysis was not performed, but we
can conclude that we have similar results for small values of $m_1$. 
In Ref.~\cite{Formaggio:2011jg} the authors assume null value for the
lightest mass $m_1$ which can be compared with the black
(dotted) curve in Fig.~\ref{sen}. For mixing angles $\sin^2
2\theta_s\sim0.1$, our exclusion is $\sim 0.6$ stronger in $\log_{10}
\Delta m_{41}^2$. This stronger exclusion can be the result of two
issues: i) we use the optimized running time in the measurement of
spectrum; ii) a different $\chi^2$ function can be used in Ref.~\cite{Formaggio:2011jg}. For the very small mixing angles
$\sin^2 2\theta_s\lesssim 0.05$, our analysis do not show the wiggling
behavior in Ref.~\cite{Formaggio:2011jg}
which is not expected in kinematical mass
measurement experiments. Generally, we have an agreement in the
limiting case of very small light mass $m_1$ with the previous
results. However, for non small $m_1$ masses, we found an interplay
between the mixing parameter $U_{e4}$ and the two free mass scales $m_1$
and $m_4$, that was not noticed in all previous analyses.

\section{\label{sec:conclusion}Conclusions}

In this work we analyzed the sensitivity of beta decay experiment
KATRIN in determining the mass scale associated with the presence of a
light sterile neutrino state. Motivation comes from the $\bar{\nu}_e$
reactor anomaly, the Gallium anomaly, the LSND and MiniBooNE
experiments which favor the presence of light sterile neutrinos that mix
with electron neutrinos, compatible with $\Delta{m}^2_{\text{SBL}}
\gtrsim 0.1 \, \text{eV}^2$ and small mixing angles $\sin^2
2\theta_s$.

For the first time we considered the effect of light mass scale $m_1$
in the determination of oscillation parameter $\Delta m_{41}^2$ in
KATRIN. We exploited a general treatment of nonzero values for the
lightest mass scale $m_1$ and the heavier mass scale $m_4$. We have
shown that varying the unknown mass scale $m_1\in[0,2]\,{\rm{eV}}$,
induce $0.2$ uncertainty in the sensitivity of KATRIN to $\log_{10}
\Delta m_{41}^2$.  However, we have shown that despite this
uncertainty, with 3 years of data-taking, KATRIN can exclude the main
part of the current allowed region in $(\sin^2 2\theta_s,\Delta
m_{41}^2)$ plane indicated by the global fit of short-baseline
oscillation experiments. Also, we have shown that for very small mixing angles $\sin^2
2\theta_s\lesssim10^{-1}$, the KATRIN experiment gives the strongest bound on the oscillation parameter $\Delta m_{41}^2$.

\begin{acknowledgments}
A.~E. and O.~L.~G.~P.  thank support from FAPESP. O.~L.~G.~P. thanks support from CAPES/Fulbright and Cosmology
Initiative of Arizona State University, where part of this work was
made, for the hospitality. We acknowledge the use of CENAPAD-SP and
CCJDR computing facilities.
\end{acknowledgments}

\end{document}